# Probabilistic image generation conditioned on an optical latent code

Yu Shao[a,b,d,†], Haiqi Gao[b,d,†], Fuyan Wu[b,d], Ziyan Zhang[b,d], Junren Wen[b,d], Yuchuan Shao[a,b,d,*], Weidong Shen[c], Chenying Yang[b,d,*]

[a] Shanghai Institute of Optics and Fine Mechanics, Chinese Academy of Sciences, Shanghai, 201800, China
[b] Hangzhou Institute for Advanced Study, University of Chinese Academy of Sciences, Hangzhou, 310024, China
[c] State key laboratory of Modern Optical Instrumentation, Department of Optical Engineering, Zhejiang University, Hangzhou, 310027, China
[d] University of Chinese Academy of Sciences, Beijing, 100049, China

[†] These authors contributed equally to this work.
*E-mail: shaoyuchuan@siom.ac.cn (Yuchuan Shao); ycheny@zju.edu.cn (Chenying Yang)

## Abstract

While diffractive optical networks compress images into compact representations, such measurements have predominantly served discriminative inference or deterministic reconstruction. Here we demonstrate probabilistic image generation conditioned solely on a 16-bit binary optical latent code (OLC) formed by a diffractive encoder and binarized camera readout. The intentionally many-to-one mapping makes each OLC specify a class of OLC-compatible images rather than a unique reconstruction target. An OLC-conditioned conditional variational autoencoder (CVAE) learns this distribution, while optical re-encoding promotes consistency between generated images and the measured OLC. Numerical characterization and proof-of-concept acquisition with a spatial light modulator–camera system establish the OLC as a physically accessible generative condition. On MNIST, samples generated from a single OLC vary in morphology while preserving class identity, and their optical-recognition accuracy reaches 81.6%, close to the 81.9% obtained from the source OLCs. These results link compact optical sensing to probabilistic synthesis of optically recognizable images.

## Introduction

Optical neural networks provide a promising platform for high-throughput and low-latency information processing by exploiting the parallelism of light propagation[1, 2]. Diffractive optical networks[3] have been demonstrated for image classification[4-6], mode transformation[7-9] and task-oriented sensing[3, 10-12], in which an optical front end preserves information relevant to a downstream task rather than forming a conventional image. Most existing systems, however, map an input directly to class-specific detector responses[4, 5] or continuous optical features[1, 2, 13]. Although these representations substantially reduce the dimensionality of the acquired data, their effective information capacity depends on the precision of optical detection and analog-to-digital conversion. A physically measurable representation with an explicit low-bit constraint is therefore desirable for interfacing optical sensing with subsequent digital processing[14, 15].

Recent optical autoencoders and generative processors have extended optical computing beyond classification[16-19]. Images have been mapped to compact optical latent spaces for transmission[17], reconstruction, denoising[16, 18] and inference[16, 20], while optoelectronic noise sources[21], scattering media[22] and diffractive processors have been employed for image generation[19, 23-26]. The latent representations in these systems, however, are predominantly continuous optical amplitudes, phases or intensities[16, 17, 19, 20]. Moreover, optical autoencoders generally seek deterministic reconstruction of a specific input, whereas optical generative models mainly use optical hardware as a decoder that transforms random or digitally prepared seeds into images[19, 21-23]. Whether an extremely compact binary code acquired directly by an optical front end can retain sufficient information to support probabilistic image generation remains largely unexplored.

At such an extreme information bottleneck, unique recovery of the input is intrinsically ill posed because many images inevitably share the same short code[14, 27, 28]. We therefore interpret a binary optical code not as an invertible description of one image but as a physical index of a set of compatible images. The decoding objective then changes from deterministic inversion to conditional distribution learning, in which different stochastic samples produce distinct images that remain consistent with the information carried by the code. This formulation converts the many-to-one nature of severe optical compression into a one-to-many generative process.

We establish a hybrid optical–electronic framework[1, 4, 29] that recasts image recovery from a severely compressed optical measurement as conditional distribution learning rather than deterministic inversion. A task-optimized diffractive encoder transforms each input image into a camera-plane intensity pattern, from which spatial integration, normalization and thresholding produce a 4×4 binary OLC. Because this 16-bit measurement is intentionally many-to-one, each OLC defines an equivalence class of images under the learned optical forward model rather than a unique reconstruction target. A conditional variational autoencoder (CVAE)[30] learns the corresponding conditional image distribution through an OLC-dependent prior and multiscale feature modulation, while stochastic sampling represents degrees of freedom that remain unresolved by the optical measurement. Re-encoding generated images through the frozen differentiable optical model further constrains them to remain consistent with the measured OLC. Numerical characterization and proof-of-concept acquisition with an Spatial Light Modulator(SLM)–camera system demonstrate that the OLC is both a learned latent representation and a physically accessible measurement. Comparisons with categorical conditioning and with local convolutional and global linear 16-bit digital representations further show that its contribution cannot be reduced to category identity or generic binary compression. The OLC therefore provides a physically grounded generative condition that links optical sensing to probabilistic image synthesis.

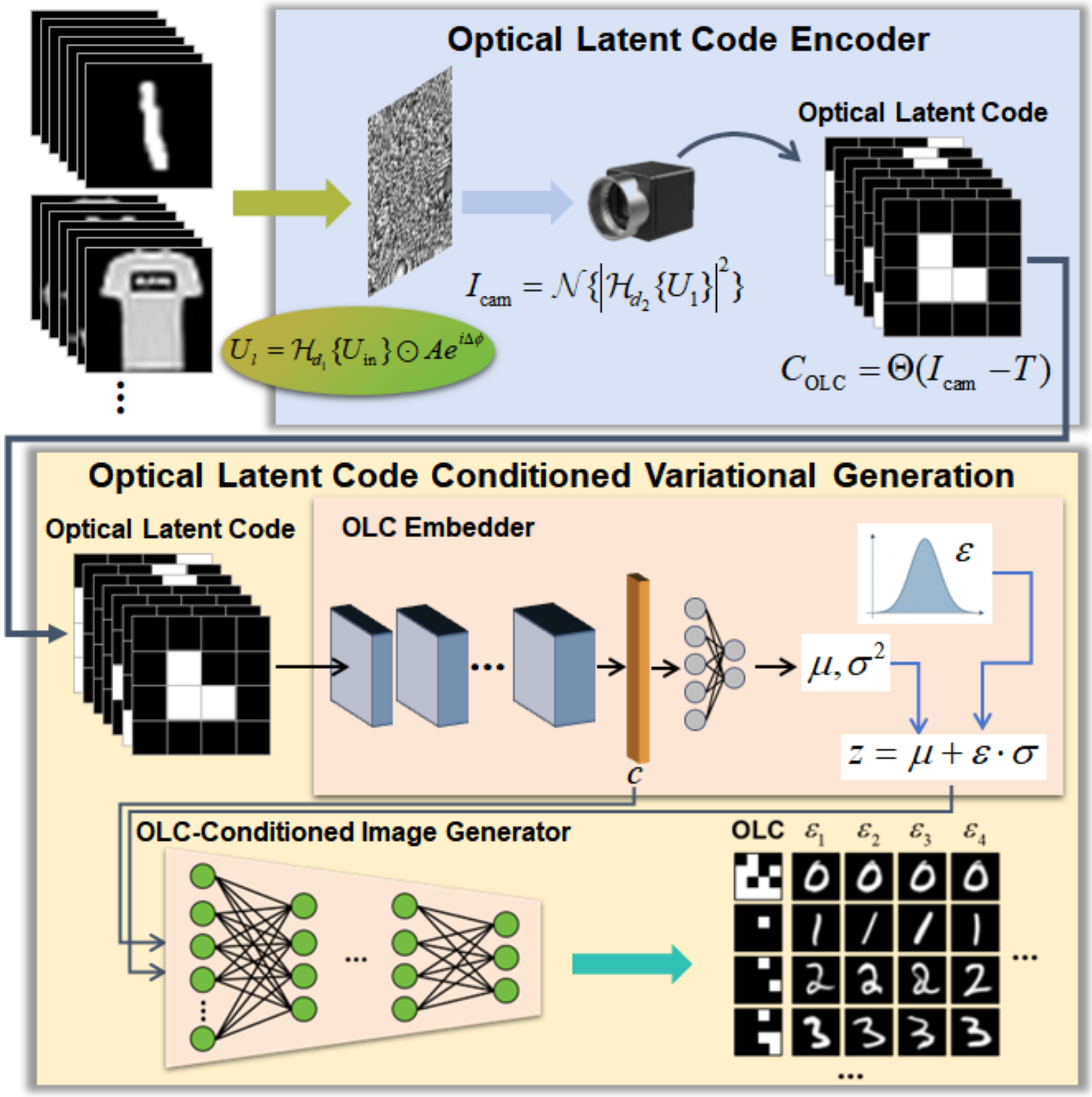

Figure 1 Conceptual framework of binary optical latent encoding and OLC-conditioned variational image generation. Top: Optical Latent Encoding (OLE). An input image is introduced as the field $U_{in}$ which propagates over a distance $d_1$, denoted $\mathcal{H}_{d_1}\{\cdot\}$, and is then modulated by the trained phase profile $Ae^{i\Delta\phi}$. Following a second propagation over $d_2$, the camera-plane intensity is spatially binned and captured as $I_{cam}$. A global threshold $T$ binarizes the normalized measurement, yielding a 4×4 optical latent code (OLC) derived from the recorded intensity. Bottom: Digital variational generation conditioned on the OLC. An embedder maps the OLC to a condition vector $c$, from which the prior modulator predicts the parameters $(\mu, \sigma^2)$ of a code-dependent Gaussian distribution. A latent vector is sampled through the reparameterization $z = \mu + \varepsilon \cdot \sigma$, where $\varepsilon$ is standard Gaussian noise. The sampled $z$ and condition vector $c$ are then supplied to the image generator. The output panel shows four samples generated from the same OLC.

## Results

### Optical latent encoding compresses images into 16-bit optical latent codes

We first examined whether a single diffractive layer could compress an image into a compact binary representation while retaining information relevant to recognition. To this end, the optical latent encoder and a diffractive recognizer were optimized jointly with a classification objective[4, 5]. Diffraction maps the input field onto a two-dimensional intensity pattern at the camera plane. Thus, although the encoder is trained using class labels, the spatial distribution of the measured intensity allows the OLC to retain part of the image structure beyond class identity. As illustrated in Fig. 2a, the 28×28 input image passes through a trainable phase modulation layer and is then recorded by the camera, and the corresponding diffraction model is described in Methods. The recorded intensity is averaged over 4×4 cells, and each normalized cell value $I_{cam}$ is binarized with a learned threshold $T$,

$$C_{\mathrm{OLC}} = \Theta(I_{\mathrm{cam}} - T) \tag{1}$$

where $\Theta(\cdot)$ is the Heaviside function. This process maps each $28 \times 28$ input image to a 16-bit OLC, reducing the nominal representation size by a factor of 392 relative to an 8-bit grayscale image. The OLC then serves as the input to the optical recognizer, which maps it to ten class-specific detector regions for classification.

The official MNIST training set was divided into 54000 images for training and 6000 images for validation and model selection, with class proportions preserved in both subsets. The official test set of 10,000 images was used only for final evaluation. To assess the information retained after compression, we first classified the OLCs using the optical recognizer, which achieved an accuracy of 81.9% on the MNIST test set. We next characterized the OLC to quantify how optical compression affected its structure and information content. As shown in Fig. 2b, the class-averaged silhouette coefficient[31] increased from 0.043 for the original images to 0.153 for the OLCs, indicating that the encoder groups images of the same class more compactly than the pixel representation.

We next quantified the many-to-one mapping introduced by the 16-bit optical compression using OLC collisions and pairwise Hamming distances. As shown in Fig. 2c and 2d, identical OLCs occurred for 16.65% of intra-class image pairs but only 0.226% of inter-class pairs, while the mean Hamming distance was 2.88 bits within classes and 4.96 bits between classes. Beyond these pairwise measures, we assessed the overall use of the 16-bit OLC space. The 10000 test images produced 1822 distinct OLCs, with an empirical entropy of 8.15 bits. The complete characterization is provided in Supplementary Table S1. The principal component analysis (PCA) and t-distributed stochastic neighbour embedding (t-SNE) projections show that the OLCs form a compact representation with partial class separation, while images sharing the same OLC coincide, as shown in Supplementary Fig. S2. The OLC therefore retains part of the within-class variation in addition to class identity while assigning many images to each OLC, which makes deterministic inversion ill posed and motivates the probabilistic generation described below.

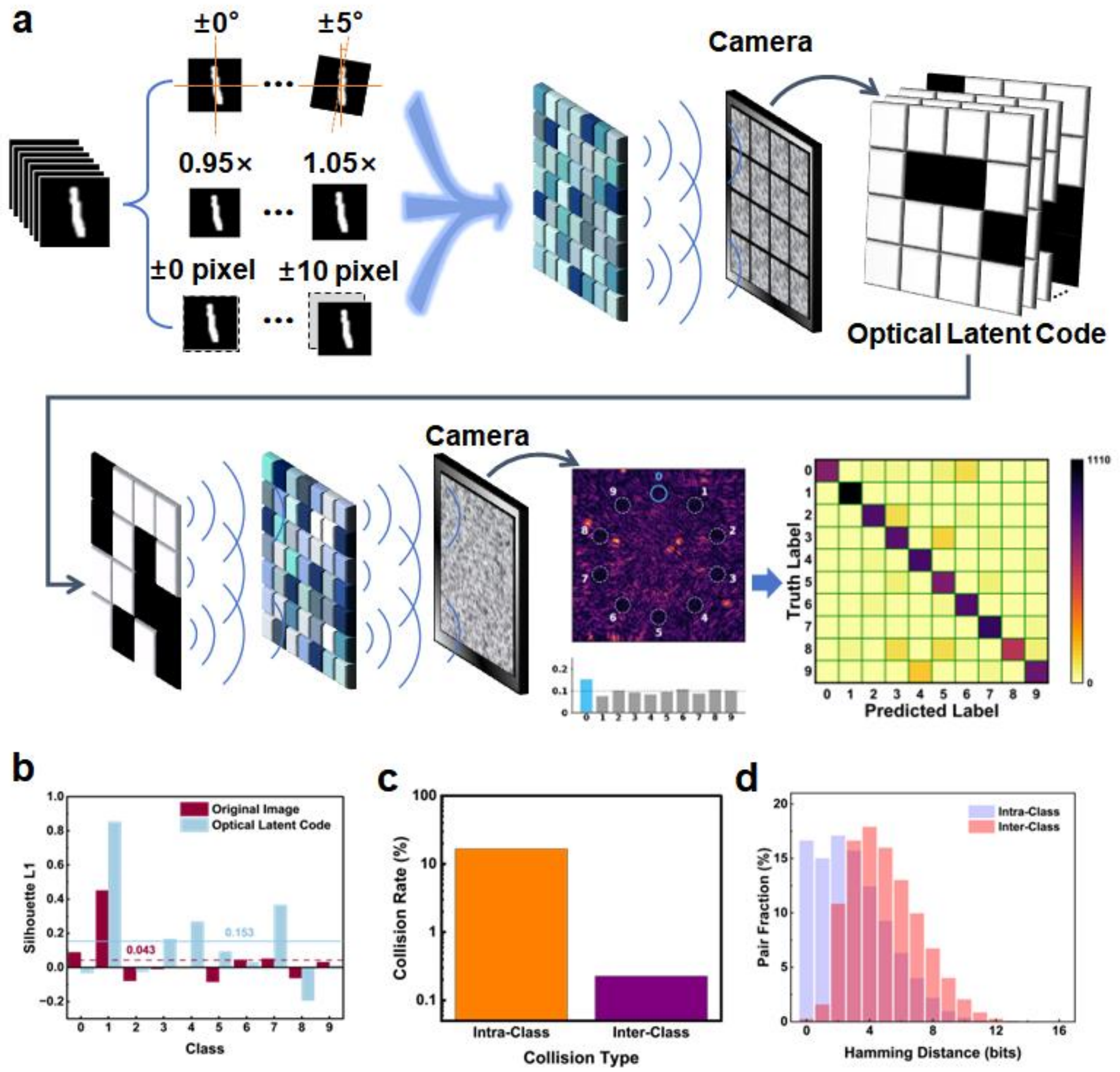

Fig. 2 Optical latent encoding and structure of the 16-bit OLC. (a) Training and inference pipeline. Input images augmented by rotation, scaling and translation are encoded by a trainable diffractive phase layer, recorded by a camera, integrated over 4 × 4 cells and thresholded into a binary optical latent code (OLC). The OLC drives the optical recognizer, a second diffractive layer whose output energy at ten detector regions gives the class. Bottom right, confusion matrix on the MNIST test set with an accuracy of 81.9%. (b) Class-resolved silhouette coefficients of the original images with L1 distance and of the OLCs with Hamming distance. Horizontal lines mark the averages over the ten classes. (c) Intra-class and inter-class collision rates on a logarithmic scale. (d) Pairwise Hamming-distance distributions for intra-class and inter-class image pairs. Panels b–d use the full MNIST test set.

## Experimental validation of optical latent-code acquisition

To verify that the OLC defined in simulation can be read out from a physical optical field, we implemented the optical latent encoding in the free-space setup shown in Fig. 3a. The input image and the trained phase profile were displayed on two reflective SLMs, and the camera intensity was integrated over 16 detection windows aligned with the cell positions defined in the simulation, as detailed in Methods. Ten test images of each class were measured for MNIST and for Fashion-MNIST, and the measured OLCs were compared with those obtained by numerical simulation of the same images.

The measured OLCs showed strong agreement with their numerical predictions, as shown in Fig. 3b. All MNIST samples reproduced the simulated OLCs exactly. For Fashion-MNIST, the measured and simulated results agreed for 95.9% of the bits, with 68 exact OLC matches. Although the measured intensity patterns in Fig. 3c show slight differences from the simulated patterns in their fine spatial distributions, the resulting binary OLCs remain highly consistent, suggesting that the cell-based binary readout reduces sensitivity to fine intensity variations and improves the stability of optical compression.

The lower agreement for Fashion-MNIST was mainly associated with active cells whose measured responses fell below the threshold. Of the 66 mismatched bits, 54 were active in simulation but inactive in the experiment. Because denser OLCs rely on more cells remaining above the threshold, they are more sensitive to reductions in the measured response. Consistent with this, the Fashion-MNIST OLCs contained 3.78 active bits on average, compared with 1.00 for the measured MNIST images. As the detection windows, background correction and threshold were calibrated on the same measurements, this experiment serves as a proof-of-concept verification rather than an assessment under independent calibration. It nevertheless confirms that the OLC can be acquired with an SLM–camera system, which provides the physical basis for the OLC-conditioned generation described next.

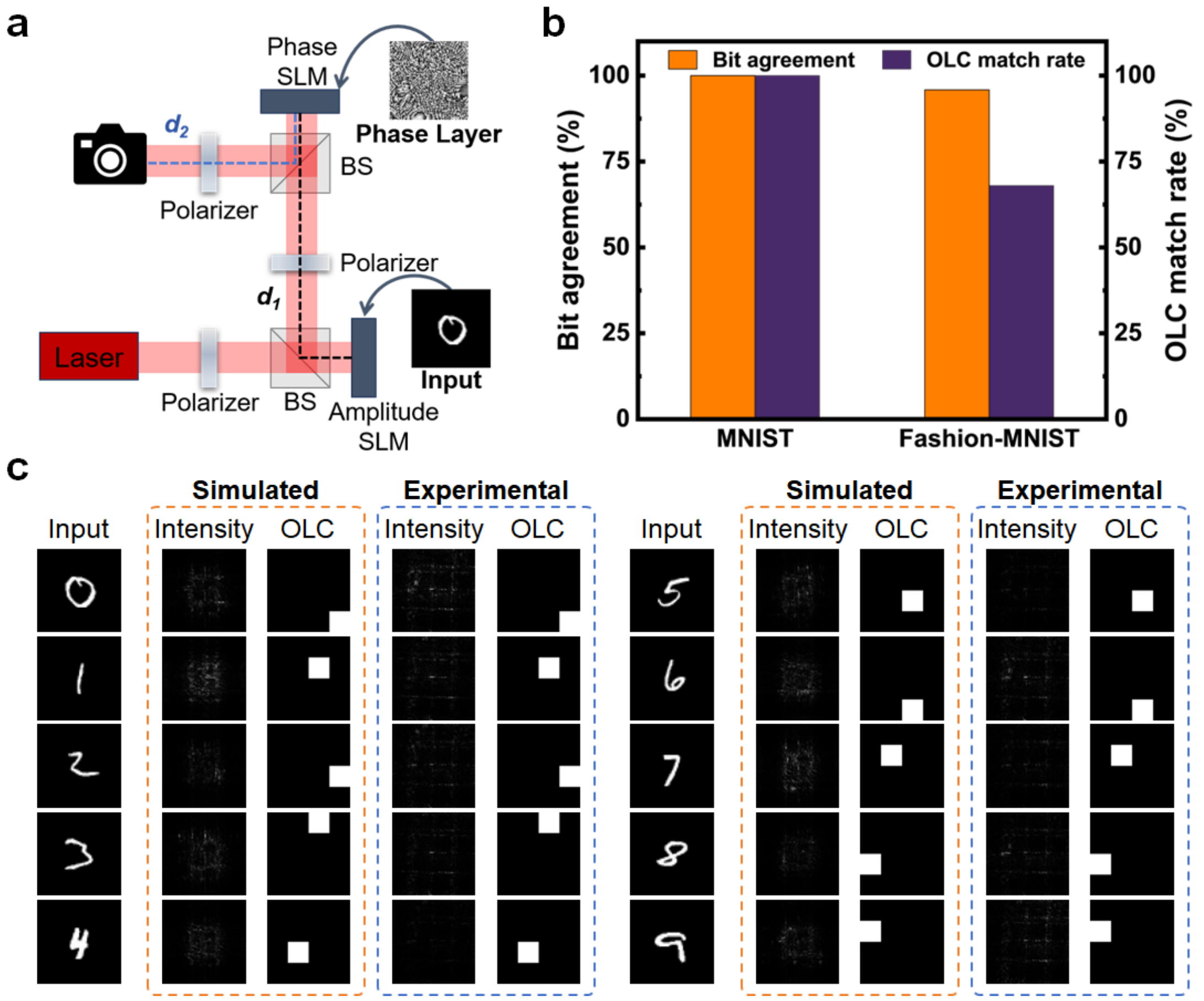

Fig. 3 Experimental validation of binary optical latent-code acquisition. (a) Schematic of the free-space optical setup. The input image and trained encoding phase profile are displayed on reflective SLMs, while polarization optics and beam splitters route the optical field to the camera after propagation over $d_1$ and $d_2$. (b) Agreement between measured and simulated OLCs for 100 MNIST and 100 Fashion-MNIST test images, given as the fraction of identical bits and the fraction of identical OLCs. Ten images of each class were selected as those with a large intensity difference between the active and the inactive cells of the OLC. (c) Representative results for the ten MNIST classes, including the input images, simulated intensity distributions and OLCs (orange outlines), and measured intensity distributions and OLCs (blue outlines). BS,

beam splitter; SLM, spatial light modulator; OLC, optical latent code.

## One-to-many generation from a single optical latent code

Since one OLC represents multiple images, we formulate decoding as conditional distribution learning rather than deterministic inversion. Following the conditional variational autoencoder (CVAE) framework[30, 32], we model the images compatible with an OLC through a latent variable whose prior is conditioned on the OLC, so that sampling this prior with different noise vectors yields different images associated with the same OLC. The OLC-conditioned CVAE, hereafter OLC-CVAE, consists of an OLC embedder, a conditional prior, an image embedder with a posterior network used only during training, and a multiscale image generator, as detailed in Supplementary Note S3. The OLC embedder maps the OLC to a condition vector $c$. Based on $c$, the conditional prior predicts the mean $\mu$ and variance $\sigma^2$ of a 16-dimensional Gaussian latent variable. The generator receives both $c$ and the latent vector as input. In addition, $c$ modulates the generator features at every spatial scale through feature-wise linear modulation(FiLM)[33], so that the influence of the OLC persists as the image is progressively synthesized.

Fig. 4a shows one test image of each digit, its OLC, the posterior reconstruction and nine images generated with the noise vectors $\varepsilon_1$~$\varepsilon_9$, which are shared by all rows. Within each row, the samples retain the digit identity of the original image while varying in stroke thickness, slant and local curvature. Across the test set, 96.92% of the generated images re-encoded to their conditioning OLC, and their optical-recognition accuracy of 81.64% was close to the 81.93% obtained from the source OLCs. The prior samples reached an average SSIM of 0.602 with respect to the original images, compared with 0.908 for posterior reconstructions, because a prior sample represents one member of the compatible set rather than the particular image from which the OLC was obtained. Their similarity to the distribution of real handwritten digits was quantified by the Fréchet Inception distance (FID), which reached 20.07 relative to the test images. Because FID is computed from Inception features that are independent of the optical model, it assesses the generated images without relying on the frozen encoder used during training. The complete evaluation is listed in Supplementary Table S3, and the evaluation protocol is described in Supplementary Note S5.

To examine whether the generator uses the within-class information carried by the OLC, we collected 1443 distinct OLCs from correctly recognized test images and decoded each of them with the same eight standard-normal noise vectors, so that differences between the generated images could be attributed to the OLC rather than to the noise. As shown in Fig. 4b, images generated from different OLCs of the same class with the same noise vector were less similar than images generated from the same OLC with different noise vectors, with the class-averaged SSIM decreasing from 0.668 to 0.598. Fig. 4c resolves this comparison by the Hamming distance between the two OLCs. The mean SSIM decreased monotonically from 0.831 to 0.438 as the Hamming distance increased from one to ten bits. These results indicate that the OLC carries within-class information that shapes generation in a graded manner. The generated image is therefore determined jointly by the measured code and the noise vector, rather than by the digit class alone.

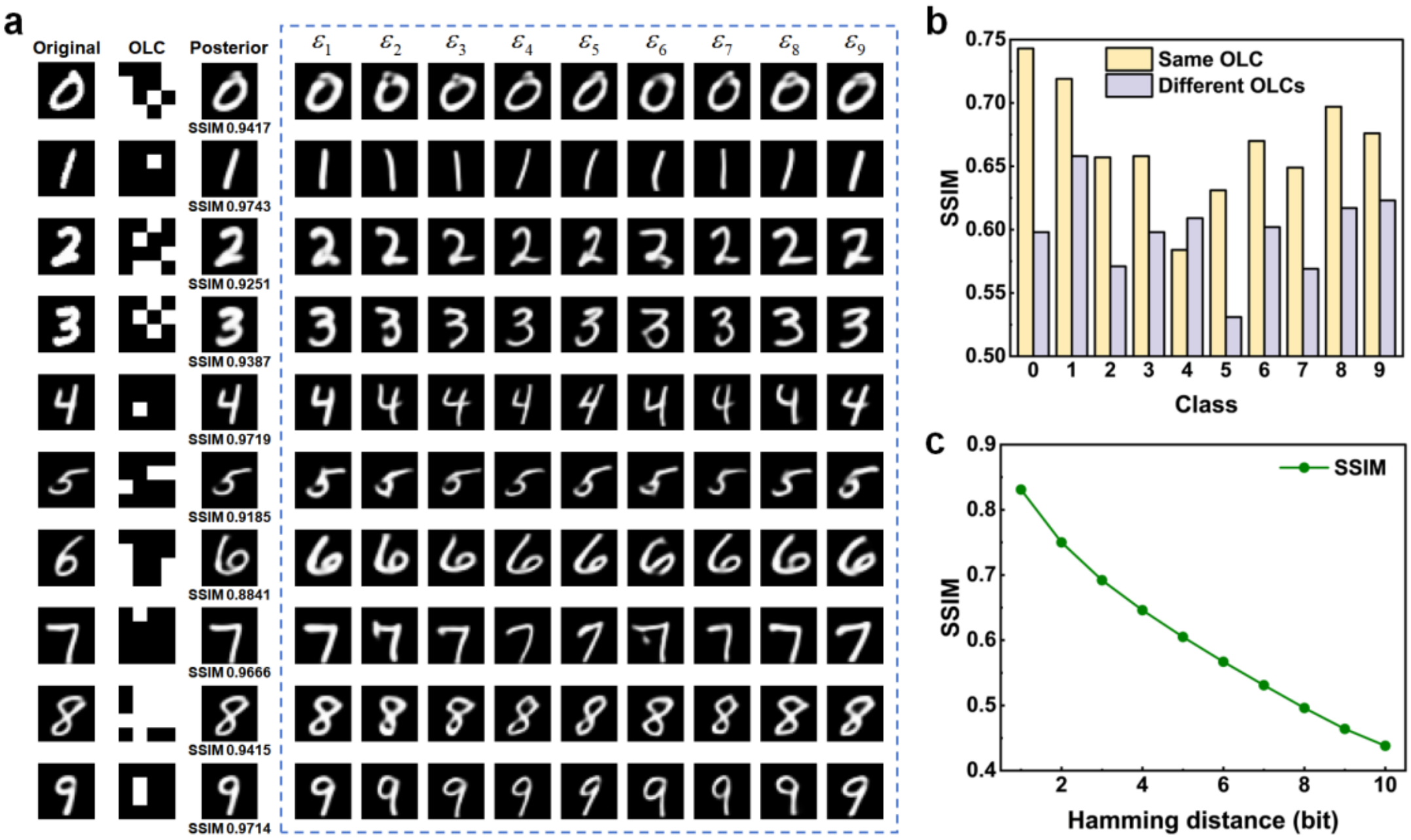

Fig. 4 One-to-many generation from a single optical latent code. (a) One test image per digit class, its 16-bit OLC, the posterior reconstruction with its SSIM and nine images sampled from the OLC-conditioned prior with the noise vectors $\varepsilon_1 \sim \varepsilon_9$, which are shared by all rows. (b) Class-resolved SSIM among images generated from the same OLC with different noise vectors and among images generated from different OLCs of the same class with the same noise vector, evaluated over 1443 distinct OLCs on the MNIST test set. (c) SSIM between images generated from two distinct OLCs of the same class with the same noise vector as a function of the Hamming distance between the OLCs, averaged over all such pairs. Distances above 10 bits, each represented by fewer than 500 pairs from at most three classes, are omitted.

**Optical latent conditioning preserves image structure and measurement consistency**

To identify the information uniquely contributed by the OLC, we compared OLC-CVAE with three reference models that share the same CVAE architecture but receive different conditions. The simplest condition is the class predicted from the source OLC by the optical recognizer, which provides Label-CVAE with the category identity but none of the within-class structure. In Conv-CVAE, every bit reads one non-overlapping $7 \times 7$ image region and encodes coarse local layout, whereas in Linear-CVAE every bit is a global projection of the image and can depend on any pixel. All four models were evaluated on the same test images with shared noise vectors. The two digital encoders were optimized jointly with their generators, and all four models were evaluated on the same test images using shared noise vectors.

Each condition imposed a characteristic trade-off among optical consistency, sample variation and source fidelity, as illustrated by the samples in Fig. 5a and quantified by the optical-recognition accuracy, the pairwise L1 diversity and the source-image SSIM in Fig. 5b, 5c and 5d, respectively. Label-CVAE generated class-typical digits that share little with the source image beyond its class, showing the lowest mean SSIM of 0.537, the largest sample diversity of 0.0971 and an accuracy of 73.94%. Linear-CVAE followed the source morphology most closely, reaching the highest SSIM of 0.677, but its samples were the least diverse at 0.0742 and its accuracy of 75.41% remained below the source level, consistent with its condition carrying the largest entropy of 11.52 bits, compared with 8.15 bits for the OLC and 7.46 bits for the convolutional code. Conv-CVAE preserved part of the local image structure yet frequently changed the digit identity, with intermediate SSIM of 0.595 and diversity of 0.0909, and it was the least recognizable optically at 43.32%. OLC-CVAE alone combined a source similarity of 0.602 and a diversity of 0.0852 with an accuracy of 81.64%, approaching the 81.93% obtained directly from the source OLCs, and it also attained the lowest FID of 20.07, compared with 21.71 for Linear-CVAE, 22.91 for Conv-CVAE and 25.18 for Label-CVAE.

These comparisons indicate that the four conditions differ in representational content, not merely in the amount of information retained. While class conditioning yields the highest generative diversity at the expense of source fidelity, convolutional conditioning preserves partial spatial structure to form clear digits but suffers from poor optical recognizability due to the absence of explicit class semantics. The proposed OLC bridges both paradigms by simultaneously conveying class labels and local structural priors. Consequently, it represents the only evaluated regime under which visually coherent samples remain robustly recognizable to the optical encoder, reaching an accuracy comparable to the source level. Crucially, because the re-encoding loss explicitly incorporates the optical encoder into OLC-CVAE training, this accuracy even surpasses the 75.41% attained by the end-to-end learned linear projection.

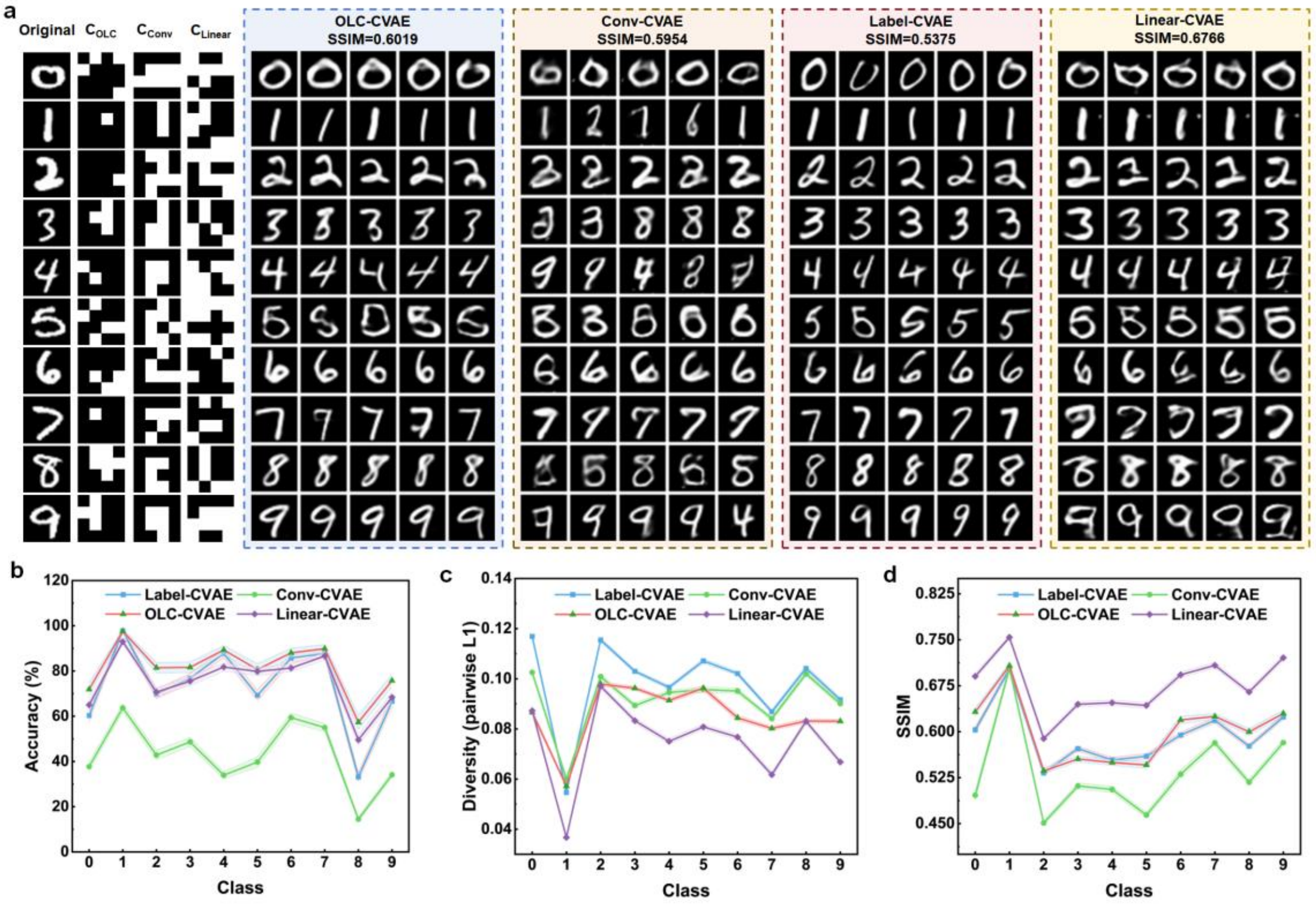

Fig. 5 Optical latent conditioning preserves image structure and OLC consistency. (a) Representative source images and their 4×4 optical, local convolutional and linear conditions, followed by five images sampled from OLC-CVAE, Conv-CVAE, Label-CVAE and Linear-CVAE. The same standard-normal noise vectors are used across the four models. Values above the generated samples denote the mean SSIM relative to the source images over the full scale MNIST test set. (b) Class-resolved optical-recognition accuracy of the source OLCs and the images generated by the four models after numerical optical re-encoding. (c) Class-resolved pairwise L1 diversity among the eight samples generated for each test image. (d) Class-resolved SSIM between generated and source images. Shaded regions denote 95% confidence intervals.

## Discussion

This work reframes an extremely compressed optical measurement as a condition for probabilistic generation rather than an invertible representation of a single image. The many-to-one encoder partitions image space into sets of inputs that share the same OLC, while the generator learns the conditional distribution associated with each set. Optical re-encoding then evaluates whether a generated image belongs to the set defined by its conditioning OLC[14, 27], rather than whether it reproduces a specific target image. In this sense, the ambiguity introduced by severe optical compression is not merely a reconstruction bottleneck, but the very basis for one-to-many generation.

The learned OLC preserves structural information beyond category identity. Within a class, varying the OLC alters the generated morphology more significantly than stochastic latent sampling under a fixed OLC, and sample similarity decreases monotonically with increasing OLC Hamming distance. Comparisons with class-label, local convolutional and linear conditioning further show that neither code length nor pixel-level similarity alone governs this behavior. Instead, these conditions partition image space differently. The OLC is distinguished by its ability to support generation while maintaining optical recognizability. The diffractive encoder therefore determines not only the information throughput of the optical bottleneck, but also the specific family of images that remains admissible after measurement. The complete generation metrics and evaluation protocol are provided in Supplementary Note S5 and Supplementary Table S3.

Designing such an optical latent code involves a fundamental trade-off between informational capacity and physical realizability. As detailed in Supplementary Note S6, increasing the OLC size preserves richer image information and improves distributional similarity, but also reduces the number of training examples available for each condition and makes exact re-encoding more demanding. Physically, smaller detection cells would further increase sensitivity to optical power, alignment and noise. Conversely, a shorter OLC is easily reproduced but partitions the image space too coarsely. The 4×4 OLC provides a balanced compromise between information capacity, generative consistency and physical implementability. The corresponding results on Fashion-MNIST suggest that this principle extends beyond handwritten digits, although richer morphology and texture remain more challenging to represent through the same binary bottleneck.

In the present demonstration, the OLC is acquired optically, whereas generation and re-encoding are performed numerically, and the optical encoder is trained with a classification objective. For complex scenes lacking discrete classes, classification accuracy no longer serves as an adequate criterion for code selection. Future implementations must adopt criteria aligned with the target distribution—such as information-theoretic constraints, perceptual metrics, or end-to-end co-training with the generator[20, 28]. A further extension is to implement the complete pipeline optically, in which all-optical compression delivers the OLC without electronic readout and all-optical generation synthesizes the images from the measured OLC. More generally, the results identify a distinct role for diffractive optics in generative imaging. Rather than reconstructing the scene, the optical front end defines a compact and physically measurable constraint on the distribution of admissible images, which establishes a route from task-specific optical inference towards OLC-consistent probabilistic generation.

## Methods

### Diffractive encoder model

Light propagation was modelled under the scalar diffraction approximation on a 1024 × 1024 grid with a sampling interval of 8 μm at a wavelength of $\lambda$ = 671 nm. Each image, normalized to [0, 1] and upsampled to 532 × 532 pixels by nearest-neighbour interpolation, was zero-padded to the full grid and used as the amplitude of the input field $U_{\text{in}}$. As illustrated in Fig. 1, the field propagates over $d_1$=0.237 m to the phase layer and over $d_2$=0.210 m to the camera. Both propagation steps were computed as follows:

$$U_1 = \mathcal{H}_{d_1}\{U_{\text{in}}\} \odot Ae^{i\Delta\phi} \tag{2}$$

$$I_{\text{cam}} = \mathcal{N}\{\left|\mathcal{H}_{d_2}\{U_1\}\right|^2\} \tag{3}$$

where $\odot$ denotes element-wise multiplication, $Ae^{i\Delta\phi}$ is the transmission of the phase layer with $A = 1$, and $\mathcal{H}_d\{\cdot\}$ denotes angular-spectrum propagation[34] over a distance $d$,

$$\mathcal{H}_d\{U\} = \mathcal{F}^{-1}\{\mathcal{F}\{U\}\exp(i2\pi d\sqrt{\lambda^{-2} - f_x^2 - f_y^2})\} \tag{4}$$

where $\mathcal{F}$ denotes the two-dimensional Fourier transform, $f_x$ and $f_y$ are the spatial frequencies.

The phase profile $\Delta\phi$ consists of 76 × 76 trainable macropixels of 7 × 7 pixels each, and its values were quantized to the 1024 levels of the phase response measured for the spatial light modulator at 671 nm. The readout operator $\mathcal{N}$ normalizes the camera intensity to its maximum, averages it over 4 × 4 equal cells and normalizes the cell values by their maximum, and the OLC is obtained from Eq. (1) with a learned threshold $T$ between 0 and 1. The 2 × 2 and 7 × 7 OLCs in Supplementary Note S6 were obtained with the corresponding numbers of cells. Gradients were passed through the phase quantization and the binarization with straight-through estimators, using a sigmoid surrogate with a temperature of 0.1 for the latter.

The optical recognizer is a second diffractive stage with the same structure and propagation distances as the encoder but an independent phase profile. The OLC, upsampled to 532 × 532 pixels, served as its input amplitude, and the energies collected by ten circular detector regions with a radius of 24 pixels, placed uniformly on a ring of radius 190 pixels, served as the class scores, the largest of which gives the predicted class.

### Training of the optical encoder and optical recognizer

MNIST[35] and Fashion-MNIST[36] were used with their official partitions of 60000 training and 10000 test images of 28 × 28 pixels. For each dataset, 10% of the training images of each class were held out as a validation set with a fixed random seed, giving 54000 training and 6000 validation images for MNIST and 54000 and 6000 images for Fashion-MNIST. MNIST was used for the main numerical encoding and generative analyses. Both MNIST and Fashion-MNIST were used in the physical acquisition experiment, and Fashion-MNIST was additionally analyzed in Supplementary Notes S2, S4 and S5. The phase profiles of the encoder and the optical recognizer, together with the binarization threshold, were optimized jointly by minimizing the cross-entropy between the softmax of the ten detector energies and the class label. During training, each input was randomly rotated by up to ±5°, scaled by a factor between 0.95 and 1.05 and translated by up to ±10 pixels on the 532 × 532 grid. To improve robustness to experimental imperfections, the normalized camera intensity was further perturbed before cell averaging by a random lateral shift of up to 5 pixels, a multiplicative gain drawn uniformly from [0.95, 1.05] and additive Gaussian noise with a standard deviation of 0.05. Augmentation and perturbations were disabled during evaluation.

Training used the Adam optimizer with a learning rate of $10^{-3}$, a batch size of 128, cosine annealing over 100 epochs and gradient clipping at a norm of 1.0, and the checkpoint with the highest validation accuracy was selected. For Fashion-MNIST, the perturbations were disabled for the first 20 epochs and increased linearly to full strength over the following 20 epochs, and only checkpoints trained with full perturbations were eligible for selection. The 2 × 2 and 7 × 7 encoders were trained in the same way as the 4 × 4 MNIST encoder.

The OLCs were characterized on the test set. A collision denotes a pair of images that produce identical OLCs, and the intra-class and inter-class collision rates are the corresponding fractions of colliding pairs among all same-class and different-class image pairs. The empirical entropy was computed as $-\sum_{o_i} p_{o_i} \log_2 p_{o_i}$ ,where $p_{o_i}$ is the relative frequency of OLC $o_i$. Silhouette coefficients were computed with the L1 distance in image space and the Hamming distance in OLC space. For each test sample $i$,

$$s_i = \frac{b_i - a_i}{\max(a_i, b_i)} \tag{5}$$

where $a_i$ is the mean distance from $i$ to the other samples of its class and $b_i$ is the smallest mean distance from $i$ to the samples of another class. The coefficients were averaged first within each class and then over the ten classes.

**Experimental implementation**

The input image was displayed on a reflective spatial light modulator (UPOLabs, HDSLM80R) operated in amplitude-modulation mode, which encoded the image in the amplitude of the incident field. After propagation over $d_1$, the field was modulated by the trained phase profile displayed on a second spatial light modulator of the same model operated in phase-only mode, and the intensity after the second propagation over $d_2$ was recorded by a camera (Hikrobot, MV-CE200-11UM). Beam splitters and polarization optics folded the free-space path to the propagation distances used in simulation and set the polarization states required for amplitude and phase modulation. Sixteen detection windows registered to the simulated cells integrated the measured intensity, and the integrated responses were background-corrected, normalized and thresholded to obtain the 4×4 OLC. Window registration, background correction and decision threshold were adjusted separately for each dataset on the same measurement batches that were subsequently evaluated. One hundred MNIST and one hundred Fashion-MNIST test images, ten from each class, were measured. Within each class, the images used in the experiment were selected according to the contrast of their simulated camera-plane intensity, so that the active and inactive detection cells of the selected OLCs were well separated from the decision threshold. Agreement between the measured and simulated OLCs was quantified by the fraction of identical bits among the 1600 measured bits of each dataset and by the fraction of the 100 OLCs whose 16 bits were all identical.

**OLC-conditioned variational generator and reference models**

We implemented a CVAE conditioned on the OLC to model the distribution of images compatible with each 4×4 binary OLC. The OLC encoder comprises a 16-channel 3×3 convolution followed by group normalization, a SiLU activation and a fully connected layer with layer normalization, producing a 128-dimensional condition vector. The image encoder applies a 3×3 convolution with a stride of 4 to produce 16×7×7 features, which are projected to 128 dimensions. The image and condition representations are concatenated and passed through two 256-unit fully connected layers to parameterize a 16-dimensional diagonal Gaussian posterior. A zero-initialized linear layer predicts the mean and log variance of the conditional prior directly from the OLC representation. The generator projects the concatenated condition and latent vectors to a 64×7×7 feature map, following by two bilinear upsampling stages yielding 32×14×14 and 16×28×28 feature representations. Feature-wise linear modulation (FiLM) residual blocks are applied at each resolution scale, and a final 3×3 convolution with sigmoid activation outputs the synthesized image. The network architecture is illustrated in Supplementary Fig. S4, with further implementation details provided in Supplementary Note S3.

Because a single OLC can be shared by images from different classes, using all associated images as targets can introduce conflicting semantic supervision. Each OLC was therefore paired with two target images of its majority class among the training images sharing that OLC; each target was reconstructed through the posterior and matched by one prior sample. Ties were assigned to the smaller class index, and sampling with replacement was used when fewer than two matching instances were available. Training optimized six objectives concurrently. Posterior reconstructions were supervised with binary cross-entropy, while the posterior was regularized against the conditional prior with a free-nats floor of 0.25 nats per latent dimension, applied to the minibatch-mean KL divergence. To constrain the coarse structure of prior samples without penalizing high-frequency stroke variations, prior samples and their target images were bilinearly downsampled to 14×14 pixels and evaluated using an $L1$ loss with a weight of 0.1. Sample diversity between pairs of prior samples was encouraged using a hinge loss with a margin of 0.05 and a weight of 0.2. To enforce physical consistency, posterior reconstructions were passed through the fixed numerical diffractive encoder and penalized when the resulting logits disagreed with the conditioning OLC under a unit logit margin. The weight of the KL divergence term was linearly scheduled to $10^{-3}$ over the first 30 epochs. The optical re-encoding loss was activated at epoch 20 and increased linearly to a weight of 0.3 over the subsequent 20 epochs.

The network was trained for 50 epochs using AdamW with a learning rate of $10^{-4}$, weight decay of $10^{-5}$, batch size of 128, and gradient norm clipping at 1.0. Checkpoints were selected using a fixed subset of 1024 validation OLCs, each decoded with eight fixed standard-normal noise vectors. Selection prioritized minimizing the mean Hamming distance between the conditioning and numerically re-encoded OLCs. Differences below 0.01 bits were treated as ties and resolved by the 14×14 prior $L_1$ error, where differences under 0.005 were similarly treated as ties. Remaining ties were resolved in favour of greater high-frequency diversity. Neither SSIM nor FID was used for checkpoint selection. The official test set was evaluated only after the best validation checkpoint had been fixed.

Three reference models were trained to benchmark optical conditioning against categorical and learned digital representations. Label-CVAE used the class predicted from the source OLC by the optical recognizer as a one-hot condition, supplying categorical labels without within-class structure. Conv-CVAE learned a 4×4 binary representation using a single 7×7 convolution with a stride of 7, mapping each bit to a non-overlapping local image patch. Linear-CVAE learned a global affine projection from the 784 image pixels to 16 binary values, allowing each bit to capture global image context. The convolutional and linear encoders were binarized using straight-through estimators and optimized end-to-end alongside their generators. All four models used the same 16-dimensional stochastic latent space and generator backbone, and the Fashion-MNIST model followed the same validation-based training and selection procedure.

**Funding**. This work is supported by National Natural Science Foundation of China (62575081), Zhejiang Provincial Natural Science Foundation of China (LHZSZ26F050003), China Postdoctoral Science Foundation (2025M780775), and Research Funds of Hangzhou Institute for Advanced Study, UCAS (2023HIAS-Y008).

**Author contributions.** Y.S. and H.G. contributed equally to this work. Y.S. developed the theoretical framework and computational model, performed the numerical simulations, analyzed the results, and wrote the original draft. H.G. designed and performed the optical experiments, conducted system calibration and data analysis, and contributed to the preparation of the manuscript. F.W., Z.Z., and J.W. assisted with the experimental implementation, data collection, and analysis. Y.C.S. and C.Y. conceived and supervised the project, provided guidance on the methodology and experimental design, and reviewed and revised the manuscript. All authors discussed the results and approved the final version of the manuscript.

**Disclosures.** The authors declare no conflicts of interest.

**Data availability.** The standard datasets used in this study (MNIST and Fashion-MNIST) are publicly available. Data underlying the results presented in this paper, including the customized detector masks and the optimized phase profiles, are available from the authors upon reasonable request.

**Supplementary Materials.** The online version contains supplementary material available at https://doi.org/XXXX.